\documentclass[prd,aps,8pt,twocolumn,preprintnumbers,amsmath,amssymb,nofootinbib,superscriptaddress,a4paper,showpacs]{revtex4-1}

 \usepackage[colorlinks=true,hyperfootnotes=true,citecolor=cyan]{hyperref}
\usepackage[english]{babel}
\usepackage{graphicx}
\usepackage{xcolor}
\usepackage{amsfonts,amsthm}
\usepackage{bm,bbm}

\newcommand{\lnm}{\ensuremath{\Lambda_{Q}}}

\newcommand{\lr}[1]{\left(#1\right)}

\newcommand{\beq}{\begin{equation}}
\newcommand{\eeq}{\end{equation}}
\newcommand{\ba}{\begin{eqnarray}}
\newcommand{\ea}{\end{eqnarray}}

\def\bs{\begin{subequations}}
\def\es{\end{subequations}}

\renewcommand{\geq}{\geqslant}

\def\cR{\mathcal{R}}
\def\cS{\mathcal{S}}

\newcommand{\tia}[1]{}

\newcounter{listcounter}

\begin{document}

\title{Born-Infeld Gravity: Constraints from Light-by-Light Scattering and an Effective Field Theory Perspective}

\author{Jose Beltr\'an Jim\'enez}
\email[]{jose.beltran@usal.es}
\affiliation{Departamento de F\'isica Fundamental and IUFFyM, Universidad de Salamanca, E-37008 Salamanca, Spain.}

\author{Adri\`a Delhom}
\email{adria.delhom@uv.es}
\affiliation{Departamento de F\'{i}sica Te\'{o}rica and IFIC, Centro Mixto Universidad de
Valencia - CSIC. Universidad de Valencia, Burjassot-46100, Valencia, Spain}

\author{Gonzalo J. Olmo}
\email{gonzalo.olmo@csic.es}
\affiliation{Departamento de F\'{i}sica Te\'{o}rica and IFIC, Centro Mixto Universidad de
Valencia - CSIC. Universidad de Valencia, Burjassot-46100, Valencia, Spain}
\affiliation{Departamento de F\'isica, Universidade Federal da
Para\'\i ba, 58051-900 Jo\~ao Pessoa, Para\'\i ba, Brazil}

\author{Emanuele Orazi} \email{orazi.emanuele@gmail.com}
\affiliation{ International Institute of Physics, Federal University of Rio Grande do Norte,
Campus Universit\'ario-Lagoa Nova, Natal-RN 59078-970, Brazil}
\affiliation{Escola de Ciencia e Tecnologia, Universidade Federal do Rio Grande do Norte, Caixa Postal 1524, Natal-RN 59078-970, Brazil}

\begin{abstract}
\noindent By using a novel technique that establishes a correspondence between general relativity and metric-affine theories based on the Ricci tensor, we are able to set stringent constraints on the free parameter of Born-Infeld gravity from the ones recently obtained for Born-Infeld electrodynamics by using light-by-light scattering data from ATLAS. 
We also discuss how these gravity theories plus matter fit within an effective field theory framework.
\end{abstract}

\maketitle

After Dirac's proposal of a relativistic theory for the electron \cite{Dirac:1928hu}, Halpern and Heisenberg \cite{Halpern:1933dya,Heisenberg:1934pza} realized that quantum effects would induce light-by-light scattering {(via 1-loop processes involving charged particles)}, which was forbidden in the classical theory. First calculations of this effect were performed  by Euler and Kockel in the low frequency limit \cite{Euler:1935uk}, and later on, a complete calculation of light-by-light scattering within QED was published by {Euler and Heisenberg and by Karplus and Neumann} \cite{Heisenberg:1935qt,Karplus:1950zz,Karplus:1950zza}. Several attempts to observe this phenomenon have been performed since then, focusing on different photon energy ranges \cite{PhysRevD.8.3813,Moulin:1996vv,Bernard:2010dx,Ishikawa:2012vs,Inada:2014srv,Yamaji:2016xws} {(see also \cite{Fouche:2016qqj} for a compilation of different set-ups that constrain non-linear electrodynamics)}. It was not until 2017 that the ATLAS Collaboration presented strong evidence for this process in ultra-peripherical heavy ion collisions at LHC  \cite{Aaboud:2017bwk}. {A recent account of the most up-to-date light-by-light scattering physics can be found in \cite{Schoeffel:2020svx}}. The results obtained in the ATLAS experiment are consistent with QED predictions, which allows to set bounds on theories of nonlinear electrodynamics (NED).  In this direction, Ellis \textit{et. al.} recently used ATLAS data to strongly constrain the mass scale of Born-Infeld (BI) electrodynamics to be $M_{\rm BI}\geq100$ GeV  \cite{Ellis:2017edi},  improving previous constraints. One of the aims of this paper is to use those data to place constraints on a Born-Infeld-like theory of gravity.

The origin of Born-Infeld electrodynamics \cite{Born:1934gh} can be traced back to some early attempts to regularize the electric field of a point charge at the charge's location and hence solve the problem of the {divergent} self-energy of point charges. {Born first attempted to tackle this issue by invoking a sort of finiteness principle and, borrowing ideas from special relativity, he suggested to non-linearly promote the Maxwell Lagrangian following a prescription similar to the passage from Galilean to special relativity. In collaboration with Infeld, they then embraced the framework of General Relativity and wrote down what is today known as Born-Infeld eletromagnetism described by the action
\begin{eqnarray}\label{BIaction}
\cS_{\rm BI}&=&M_{\rm BI}^4\int \text{d}^4x\left[\sqrt{-g}-\sqrt{\left\vert\det\left(g_{\mu\nu}+\frac{1}{M_{\rm BI}^2}F_{\mu\nu}\right)\right\vert}\;\right]\nonumber\\
\nonumber&=&M_{\rm BI}^4\int \text{d}^4x\sqrt{-g}\lr{1-\sqrt{1+\frac{F^{\mu\nu}F_{\mu\nu}}{2M_{\rm BI}^4}-\frac{(F^{\mu\nu} \tilde F_{\mu\nu})^2}{16M_{\rm BI}^8}}} \;,
\end{eqnarray}
 }
 where $M_{\rm BI}$ denotes the energy scale where modifications to QED become non-perturbative {or, equivalently, at electromagnetic  fields $\vert \vec{E}\vert,\vert \vec{B}\vert\sim\mathcal{O}(M_{\rm BI}^2)$ deviations  from Maxwell's electromagnetism become non-perturbatively relevant}. The interest of the community in BI actions was boosted by Fradkin and Tseytlin  \cite{Fradkin:1985qd}, who showed that actions of the Born-Infeld type arise in different scenarios related to M-theory. 

{The many appealing properties found in BI electromagnetism (see e.g. \cite{Ketov:2001dq} and  references therein) motivated to seek for similar extensions in the gravitational sector. A first incursion on this path was carried out by  Deser and Gibbons \cite{Deser:1998rj}, although the seemingly unavoidable presence of ghosts in these extensions detracted substantially from their phenomenological interest. It was later realised that the Palatini formulation of Born-Infeld-like gravity theories seemed less pathological  \cite{Vollick:2003qp} and these efforts eventually crystalised in a combination of the old Eddington purely affine gravity with Born-Infeld ideas that is now widely known as Eddington-inspired-Born-Infeld (EiBI) and whose action reads \cite{Banados:2010ix}}  
\beq\label{EiBIaction}
\cS_{\rm EiBI}=\pm\Lambda_Q^4\int \text{d}^4x\lr{\sqrt{-\left|g_{\mu\nu}\pm\frac{\mathcal{R}_{(\mu\nu)}(\Gamma)}{\Lambda_{\rm EiBI}^2}\right|}-\sqrt{-|g_{\mu\nu}|}} \;,
\eeq
where $\Lambda_{Q}$ is a high energy scale which can be factorized as $\Lambda_Q\equiv\sqrt{M_{\rm Pl}\Lambda_{\rm EiBI}}$ so that {GR with the standard value of the} Newton's constant is recovered in the weak curvature regime. Then the model has only one free parameter which can be represented by either $\Lambda_{\rm EiBI}$ or $\Lambda_Q$. {Although $\Lambda_{\rm EiBI}$ is the scale that suppresses the higher order curvature corrections, $\Lambda_Q$ might be endowed with a more physically transparent meaning since this is the scale that suppresses the associated interactions in the matter sector. In this respect, we can make an educated guess for a constraint on this scale from the absence of anomalous scattering amplitudes at LHC so we can impose  $\Lambda_Q$ to be larger than the scales probed at LHC, i.e., $\Lambda_Q\gtrsim 1 - 10$ TeV. Of course, the precise constraint depends on the particular process we look at, but the quoted limit is a natural lower bound to expect.} EiBI gravity has been extensively studied in recent years in cosmology, astrophysics, and in compact objects scenarios, where it exhibits a remarkable ability to yield nonsingular solutions and  interesting phenomenology. {For a review on the different approaches to Born-Infeld gravity and their phenomenological applications we refer to \cite{BeltranJimenez:2017doy}.} 

The dependence of the Ricci tensor on the connection ${\Gamma^\alpha}_{\mu\nu}$ has been written explicitly to emphasize that it is not assumed to be {\it a priori} related to the metric, but rather that metric and connection are naturally regarded as equally fundamental and independent geometrical objects (the so called metric-affine formulation). {The fact that only the symmetric part of the Ricci tensor appears in the action guarantees that the field equations for the connection are algebraic and, furthermore, it can be solved as the Levi-Civita connection of a metric that can be expressed in terms of $g_{\mu\nu}$ and the matter fields. In other words,} the connection {plays the role of} an auxiliary field that can be {integrated out and whose effect is the modification of the interactions in the matter sector}. As a matter of fact, a field redefinition of the metric allows to build an Einstein frame representation of the theory in which the matter fields are minimally coupled to the redefined (or Einstein-frame) metric but the interactions are modified. {We will discuss below how the same procedure applies to a much larger class of theories, so this property is not specific of the EiBI Lagrangian. It is crucial to mention that going to Einstein frame within these theories does not require the introduction of additional degrees of freedom because we only need to solve for the connection that is an auxiliary field. This is radically different from other theories like metric $f(R)$ where the space of propagating fields is enlarged with respect to GR. This is the reason why in the EiBI theory the whole effect of going to the Einstein frame simply consists in the generation of interactions in the matter sector for the fields that were already present in that sector in the original frame.} 

By construction, the {matter Lagrangian in the Einstein frame} retains the {physical symmetries of the original theory so that the same conserved Noether charges and degrees of freedom remain. The construction of the Einstein frame requires solving an equation for $g_{\mu\nu}$ in terms of the energy-momentum tensor of the matter sector so, as long as the energy-momentum tensor preserves the symmetries, these will remain. This is generally expected to be the case for internal symmetries that do not involve spacetime transformations. A  potential problem with gauge symmetries could arise for spin-2 fields because, as it is well-known, no gauge (diffs) invariant and Lorentz covariant energy-momentum tensor exists for a massless spin-2 field. In this case, integrating out the connection could lead to pathologies as the appearance of (ghostly) new degrees of freedom. Of course, these degrees of freedom were already latent in the theory before integrating out the connection. In any case, since having more than two massless spin-2 fields is already pathological, this should not be a cause of problems. On the other hand, it may occur that some properties like e.g. dualities are lost or might be appropriately re-formulated.} Remarkably, in the particular case of EiBI gravity coupled to Maxwell electrodynamics, it was recently shown that the Einstein-frame matter action exactly coincides with BI NED\footnote{{This is strictly true if we pick the negative sign in the gravitational   Lagrangian \eqref{EiBIaction}, otherwise the sign of the $\mathcal{O}(\beta^{-2})$ term in \eqref{BIaction} would be negative.} If the EiBI Lagrangian is written as in \cite{Delhom:2019zrb}, this choice corresponds to $\epsilon<0$.}, provided that one identifies $\Lambda_{Q}=\sqrt[\leftroot{-3}\uproot{3}4]{2} M_{BI}$ \cite{Delhom:2019zrb}. This correspondence {is precisely the one that we will exploit in the following} to set constraints on EiBI gravity {from the observed bounds on BI electromagnetism}. \\

In order to obtain the bound $M_{\rm BI}\gtrsim 100$ GeV, it was implicitly assumed in \cite{Ellis:2017edi} that the gravitational sector was described by GR, with the matter sector represented by the BI nonlinear modification of Maxwell's electrodynamics. An apparently different approach would be to take the gravity side as described by the EiBI action \eqref{EiBIaction} but keeping in the matter sector the standard Maxwell theory. However, {as explained above}, given that a field redefinition  transforms one theory into the other, the bound found in \cite{Ellis:2017edi} for the matter sector can be reinterpreted so as to set a bound on the EiBI gravitational sector. Accordingly, and using the results of \cite{Ellis:2017edi}, we find that light-by-light scattering data from ultraperipherical Pb-Pb collisions in ATLAS sets a conservative constraint to the EiBI gravity free parameter of
\begin{align}\label{constraint}
\Lambda_{Q}\gtrsim 120 \text{ GeV}
\end{align}
This constraint is similar to the most stringent constraints to EiBI up to date, although it is obtained in a more precise way\footnote{Notice that, although the constraint is only strictly valid for the negative sign in \eqref{EiBIaction}, corrections to QED of the same order of magnitude are expected for the positive sign ($\epsilon>0$ in  \cite{Delhom:2019zrb}), thus a similar bound would be obtained in this case. {Let us mention however that the sign of this parameter could be further constrained from positivity bounds so as to embed the theory into a local Lorentz invariant UV theory as nicely discussed in e.g. \cite{Adams:2006sv}. In this respect, it is interesting to note that the UV string theory embedding of BI electromagnetism can be related to the absence of superluminal propagation around any background \cite{Gibbons:2000xe}.}}. Concretely it is of the same order of magnitude as the one obtained in \cite{Delhom:2019wir} from Compton scattering data, and one order of magnitude lower than that obtained in \cite{Latorre:2017uve} from Bhabha scattering data. However, notice that both constraints were obtained by using a perturbative version of the mapping in powers of $1/\lnm$, whereas here we are using the full non-perturbative relation between the EiBi and Einstein frames of the theory. To compare with other existing constraints to EiBI {in the literature}, it is useful to use the parameter $|\kappa|\equiv2c^7\hbar^3\Lambda_Q^{-4}$ {so the bound  \eqref{constraint} translates as}
\begin{equation}
|\kappa|\lesssim 4\times10^{-24} \text{m}^5 \text{kg}^{-1}\text{s}^{-2}.
\end{equation}
Other constraints to EiBI gravity from nuclear physics, astrophysics, or gravitational wave data are several orders of magnitude weaker \cite{Avelino:2012ge,Avelino:2012qe,Jana:2017ost,Avelino:2019esh}. {As well, we note that this constraint gives an order of magnitude estimate of the constraints that would be found for generic RBG theories coupled to Maxwell electromagnetism, see {\it e.g.} \cite{Latorre:2017uve,Delhom:2019wir}.}\\

{\textbf{EiBI as a Ricci-Based Gravity theory}}

 Let us now turn our attention to the implications that the correspondence between theories that we used above may have for other theories of gravity. The EiBI theory is a member of a larger family of theories known as Ricci-Based Gravity theories (RBGs), which are characterised by an action of the form \cite{BeltranJimenez:2017doy,Afonso:2017bxr}
\beq\label{RBGaction}
\cS_{\rm RBG}=\int \text{d}^4x \sqrt{-|g|}F(g^{\mu\nu},\cR_{(\mu\nu)}(\Gamma)) +\cS_{\rm m}(g_{\mu\nu},\Psi_m) \ ,
\eeq
where $F$ is any analytic function of the metric and the symmetrized Ricci tensor, and $\cS_{\rm m}$ represents the matter action, in which the generic matter fields $\Psi_m$ are minimally coupled to the metric $g_{\mu\nu}$.\footnote{{The assumption of minimally coupled fields is not fundamental for our argument here and could be readily included as in e.g. \cite{Afonso:2017bxr}. We prefer to stick to the minimally coupled case for clarity of the exposition.}} In RBGs only the symmetric part of the Ricci tensor is considered  because the antisymmetric part is not invariant under projective transformations, ${\Gamma^\alpha}_{\mu\nu}\to {\Gamma^\alpha}_{\mu\nu}+\xi_\mu\delta^\alpha{}_\nu$, and an explicit breaking of projective symmetry leads to the propagation of unstable degrees of freedom in arbitrary backgrounds \cite{BeltranJimenez:2019acz,Jimenez:2020dpn}. 

{As in the EiBI theory, it is possible to show that the connection enters as an auxiliary field in all RBGs and they also admit an Einstein-frame representation where the gravitational sector is described by standard GR for the redefined metric $q_{\mu\nu}$ and integrating the connection out results in additional interactions in the matter sector. In this frame, matter is minimally coupled to the}  Einstein-frame metric $q_{\mu\nu}$ (see \cite{BeltranJimenez:2017doy,Afonso:2018bpv,Afonso:2018hyj,Afonso:2018mxn,Delhom:2019zrb} for further details). This redefined metric is associated to $\partial F/\partial\mathcal{R}_{(\mu\nu)}$, and its relation with $g_{\mu\nu}$ can be encoded into a {\it deformation matrix} $\hat{\Omega}$ defined via the relation
\begin{equation}\label{Relationmetric}
q_{\mu\alpha}(\Omega^{-1})^\alpha{}_\nu=g_{\mu\nu}.
\end{equation}
When the field equations are satisfied, $(\Omega^{-1})^\alpha{}_\nu$ can be written as a (model-dependent) {algebraic} function of the matter fields and one of the metrics  \cite{BeltranJimenez:2017doy,Delhom:2019zrb}, and it can be expanded in a series of the form\footnote{Here we refer to the RBGs that have GR as a low energy limit and which are characterised by a single energy scale $\Lambda_Q$ at which deviations from GR become non-perturbative. Theories characterized by more than one energy scale could be included in a straightforward manner, although they would unnecessarily obscure the argument.}
\begin{equation}\label{ExpansionOm}
(\Omega^{-1})^\mu{}_\nu=\delta^\mu{}_\nu+\sum_{n=1}^{\infty}\sum_{i_n=0}^3\frac{\beta_{(n,i_n)}}{\Lambda_Q^{4n}}(T^{(n,i_n)})^\mu{}_\nu,
\end{equation}
where each $(T^{(n,i_n)})^\mu{}_\nu$ stands for a tensor structure that can be built with $n$ powers of the stress energy tensor. 
{At each order $n$, the sum over $i_n$ runs up to three because the Cayley-Hamilton theorem guarantees that a given matrix is always a root of its characteristic polynomial. In other words, the vector space of $D$-dimensional matrices that are analytical functions of some given matrix has, at most, dimension $D$ and, in our case, we can take the first $(D-1)$ powers of $T^\mu{}_\nu$ as a basis of this space.\footnote{{Notice that the vector space spanned by the first $(D-1)$ powers of $T^\mu{}_\nu$ does not need to have dimension $D$, but its dimension will be determined by the number of linearly independent powers. In other words, the space of matrices that are analytical functions of $T^\mu{}_\nu$ has dimension $\text{rank}(\{\hat{T}^i\}\vert_{i=0,\cdots D-1}$.}} Strictly speaking, the relation \eqref{ExpansionOm} holds for the (likely more physical) branch of solutions that reduce to GR in the low energy limit. This can be important because the deformation matrix satisfies a non-linear equation in terms of the energy-momentum tensor. Thus, besides the solution \eqref{ExpansionOm} that could be obtained by imposing Lorentz covariance, there could be other branches with spontaneous symmetry breaking as discussed in \cite{Jimenez:2020iok}. However, these non-standard branches are seemingly pathological and, since we want to recover GR at low energies, \eqref{ExpansionOm} is the relevant series expansion for the solution.}

The mapping into GR for an arbitrary RBG (minimally) coupled to scalar fields, anisotropic fluids, and massless spin-1 fields has been developed in \cite{Afonso:2018bpv,Afonso:2018hyj,Afonso:2018mxn,Delhom:2019zrb}, and  also a few examples in which the full non-perturbative solution of the mapping equations can be obtained have been worked out \cite{Delhom:2019zrb,Afonso:2019fzv,Guerrero:2020azx,Shao:2020weq,Olmo:2020fnk}. For fermions only a perturbative solution of the mapping is available so far \cite{Latorre:2017uve,Delhom:2019wir} though, in principle, nothing  prevents the existence of an Einstein frame with an additional four-fermion interaction as the one appearing in the Einstein-Cartan-Sciamma-Kibble theory \cite{Kibble:1961ba,Sciama:1964wt,Hehl:1976kj}. {From the EFT viewpoint, the perturbative series is in fact the relevant regime.} 


It is important to note that extensions of the RBG family (\ref{RBGaction}) involving the Riemann tensor and its derivatives can also be considered, leading to a broader spectrum of theories. Such extensions {generically} propagate a massless spin-2 mode encoded in some $q_{\mu\nu}$ metric plus (probably) other degrees of freedom that are not present in GR. Our focus here on the RBG sector is justified because a crucial part of the action of such theories will be of the RBG type and may be constrained well before those new degrees of freedom may become observable. {Furthermore, the most general framework involving arbitrary powers of the Riemann tensor has been shown to be generically plagued by pathologies \cite{BeltranJimenez:2019acz,Jimenez:2020dpn}.}\\

\textbf{Effective matter sector in RBGs}

 The {equivalence} between RBG's and GR is particularly relevant when the matter sector is {interpreted within the realm of} an effective field theory (EFT) because constraints on nonlinearities of the matter sector can be used to place constraints on RBGs, in much the same way as we did above with the EiBI model. 
In the EFT framework (see {\it e.g.}\cite{Pich:1998xt}) one intends to give a description of the phenomena occurring below some energy scale $\Lambda$, which could take into account small corrections with origin in the UV \footnote{Here UV/low-energy refers to energies greater/lower than the EFT scale $\Lambda$. By construction, the EFT will only be physically meaningful at scales below $\Lambda$, typically breaking unitarity above this scale.} regime without actually knowing the details of the underlying UV theory. This is done by  1) identifying the set of fields that describe the spectrum at low energies and choosing a set of symmetries that should be satisfied in the UV, and 2) using those fields to build the most general Lagrangian consistent with those symmetries. 

Let us sketch how to build such Lagrangian. Assume that we have a given set of asymptotic states described by the matter fields $\Psi_i$ whose dynamics satisfies certain symmetries. This matter sector defines an EFT Lagrangian by $\mathcal{L}_{\rm{eff}}=\mathcal{L}_0+\mathcal{L}_{\rm eff}^{\rm d>4}$ where $\mathcal{L}_0$ contains the  relevant and marginal operators and $\mathcal{L}_{\rm eff}^{\rm d>4}$ is an infinite tower of all the possible irrelevant operators that can be built upon the $\Psi_i$' s and which respect the given symmetries. As is well known, the set of dimension-$d$ operators of a given quantum field theory forms a vector space $\mathcal{A}_d$. Furthermore two operators are said to be {\it equivalent}, in the sense that they contribute equally to physical observables, if they differ (up to a total derivative) by an on-shell constraint\footnote{This is an operator proportional to the {lower order} field equations, which by definition vanish on-shell see {\it e.g.} \cite{Weinberg:1995mt,Weinberg:1996kr}}. $\mathcal{A}_d$ can be split into the equivalence classes defined by this relation, and it suffices to consider one operator of each equivalent class to have a basis of $\mathcal{A}_d$ (see {\it e.g.} \cite{Einhorn:2013kja}). Thus, without loss of generality we can write	
\begin{equation}\label{EFTLag}
\mathcal{L}_{\rm eff}^{\rm d>4}=\sum_{n=5}^\infty\sum_{i_n}\frac{\alpha_{(n,i_n)}}{\Lambda^{n}}\mathcal{O}_{(n,i_n)},
\end{equation}
where the set $\{\mathcal{O}_{(n,i_n)}\}_{i_n}$ is a basis of $\mathcal{A}_n$ and $i_n$ runs from 1 to dim$\{\mathcal{A}_n\}$ for each $n$ in the above expression. The dimensionless constant $\alpha_{(p,q)}$ is the Wilson coefficient of the operator $\mathcal{O}_{(p,q)}$. Assuming {naturalness}, the Wilson coefficients will be of $\mathcal{O}(1)$ and the EFT defined by \eqref{EFTLag} will be generically valid at energies below $\Lambda$. {In this respect, it is interesting to notice that the violation of tree-level unitarity does not imply the necessity of including new physics at that precise scale as shown for instance with the self-healing mechanism discussed in \cite{Aydemir:2012nz}.} \\
 
Let us now couple the EFT described by $\mathcal{L}_{\rm eff}$ to a gravitational sector given by a particular RBG theory. {This might appear as an arbitrary construction because the EFT philosophy should also be employed in the construction of the gravity sector. However, there is nothing a priori inconsistent with considering an RBG gravity sector and our interest here is precisely to discuss how these gravity theories can be fit into an EFT framework.} Since the theory can be mapped into GR coupled to a matter sector with the same degrees of freedom and symmetries as the original one, the mapped EFT retains its structure, {\it i.e.}, the basis of operators of the original EFT is still a basis of the mapped EFT. Thus, the `new' operators that appear after the mapping were already present in the original matter Lagrangian. 

{Before proceeding further, it may be in order to digress a bit here again on the preservation of the symmetries when going to the Einstein frame. In the EFT matter sector there will be gauge symmetries that have to do with massless particles and are fundamental for the correct number of degrees of freedom. We do not expect the Einstein frame formulation of the theory to change this because the energy-momentum tensor of the gauge fields will also be gauge invariant (with the exception of massless spin-2 particles that we have already addressed above). Regarding global symmetries, these typically arise as accidental symmetries of the low energy theory but can be broken by higher dimension operators. Thus, when going to the Einstein frame in the RBGs no new operators will be generated. }
  
{In view of the above discussion and} since the original EFT matter Lagrangian was built with an arbitrary linear combination (defined by the $\alpha_{(n,i_n)}$) of the $\{\mathcal{O}_{(n,i_n)}\}_{i_n}$, which provide a full basis of the space of operators of the theory, {it should be clear that} the effect of the non-linearities introduced in the matter sector after the mapping can be reabsorbed by a redefinition of the Wilson coefficients\footnote{Note that by the structure of \eqref{ExpansionOm}, the new operators that will enter the Lagrangian after the mapping are of mass dimension $4n$, and therefore only the Wilson coefficients corresponding to $4n$-dimensional operators will be non-trivially changed.} $\alpha_{(n,i_n)}\mapsto\tilde{\alpha}_{(n,i_n)}$ which schematically looks like 
\begin{equation}\label{redefWilson}
\tilde{\alpha}_{(4n,i_n)}=\alpha_{(4n,i_n)}+\sum_{i_k}\beta_{(n-4,i_k)}\lr{\frac{\Lambda}{\Lambda_Q}}^{n-4}.
\end{equation}
 where the $\beta_{(n,i_n)}$ are (up to a constant factor) those appearing in \eqref{ExpansionOm} and $\Lambda_Q$ is the high-energy scale that parametrises departures from GR within the given RBG model. Since by construction the Wilson coefficients are arbitrary in an EFT\footnote{Of course, they can be constrained by experiments, thus ruling out regions of parameter space. But regarding the theoretical construction of the EFT, these are arbitrary coefficients}, this redefinition is not relevant, and the matter sector is described by the same EFT before and after the mapping. There is a subtlety that can arise here: If the scale $\Lambda_Q$ is much below the EFT scale, then the redefinitions in \eqref{redefWilson} imply that either the $\alpha_{(4n,i_n)}$ or the $\tilde{\alpha}_{(4n,i_n)}$ will typically be far from unity by an amount of the same order, which could restrict the range of validity of the EFT to energies below $\Lambda_Q$.\\

{\bf Discussion and conclusions}.  

As a consequence of the above discussions we arrive at the following conclusion: {the predictions of a given RBG coupled to a given matter sector will differ, in general, from those of GR coupled to the same matter sector (when the connection is integrated out), but if the matter sector coupled to the RBG is an EFT its predictions will be indistinguishable from those of GR coupled to an EFT.} This has been illustrated {in the first part of this paper} by the coupling of EiBI gravity to Maxwell, which is equivalent to the coupling of GR to Born-Infeld electromagnetism rather than of GR to Maxwell. In the EFT framework, this is stated by saying that the operators pertaining to the RBG sub-class of metric-affine actions are redundant when one considers all the allowed operators in the matter sector. Although other metric-affine models can also be shown to be equivalent to GR, it is still an open question whether there is any other disjoint class of metric-affine operators which are redundant in the sense described above. 

If the EFT approach is extended to the gravitational interactions, it is straightforward to see that the EFT for a general metric-affine sector cannot be reduced to an RBG by imposing any symmetries involving the metric and/or affine connection, even at the lowest order. In fact, RBGs only propagate a massless spin-2 mode, and the most general diffeomorphism and projective invariant action propagates extra degrees of freedom already at quadratic order, some of which are likely to be unstable unless fine-tuned models are considered \cite{BeltranJimenez:2019acz,Jimenez:2020dpn,Percacci:2019hxn}. Furthermore, it is easy to see that by imposing extra symmetries to this action one would forbid operators that lie within the RBG class. Hence, it is difficult to envision how a general metric-affine EFT could be reduced to an RBG theory by means of symmetries even at the lowest order.  Nonetheless, this does not preclude RBGs from being regarded as physically sound effective theories for perturbative phenomena at energy scales $E<\lnm$ {but then these effects would already be there from the irrelevant operators of the matter sector}. 

From a non-perturbative perspective, there are RBGs able to remove cosmological and black hole singularities
 at the classical level, restoring geodesic completeness via the emergence of wormholes or cosmological bounces. However, quantum corrections in the matter sector of those theories are likely to strongly backreact onto those backgrounds, potentially rendering them as unphysical \cite{BeltranJimenez:2017uwv}. In this regard, it is important to recall that there are effective theories with irrelevant operators that satisfy non-renormalization theorems. {This property is important at a phenomenological level because it allows to have non-perturbative classical effects from non-renormalisable operators while maintaining the quantum corrections under control. The paradigmatic example is of course General Relativity where the Planck mass in front of the Einstein-Hilbert term does not get renormalised by graviton loops.\footnote{{There is a simple argument that explains why this is the case. If we consider the so-called $\Gamma\Gamma$ Einstein form for the GR action (i.e. the Einstein-Hilbert Lagrangian deprived of the total derivative term), then diffeomorphisms are only realised up to a boundary term. Since Feynman diagrams realise the symmetries in an exact form, all loops can only generate quantum corrections to the higher order terms.}} It receives quantum corrections from matter loops, but these are typically $\mathcal{O}(m^2/M_{\rm pl}^2)\lesssim 10^{-30}$ for the standard model particles so GR is actually an excellent quantum EFT (see e.g. \cite{Donoghue:1995cz,Dobado:1997jx,Burgess:2003jk})\footnote{{We cannot resist referring to Weinberg's words on the topic in \cite{Weinberg:2009bg}.}}. Of course, there is the problem of the cosmological constant, but this is a naturalness problem rather than a breakdown of the EFT. }

{Similarly, theories like non-linear electrodynamics, $K$-essence, or Galileon theories exhibit analogous properties for their quantum corrections. Thus, if we take RBG theories and the matter sector is composed by e.g. a massless scalar field solely, the Einstein frame version of the theory after integrating out the connection will give rise to a $K$-essence model with a Lagrangian of the form $\mathcal{L}=\Lambda^4_\phi K(\partial_\mu\phi\partial^\mu\phi/M^4)$, with $\Lambda$ and $M$ some mass scales. The non-perturbative regime of the original RBG theory can then be mapped into a phenomenological effect arising from non-renormalisable operators in $K$ where $M$ corresponds to the scale of non-linearities and $\Lambda$ is parameterically given by $\Lambda\simeq\sqrt{M_{\rm pl} M}$. The structure of these scalar theories, in particular their shift symmetry, guarantees that the quantum corrections will come in with derivatives of the Lagrangian $\sim \partial K$ so it is plausible to have a regime where $\partial_\mu\phi\partial^\mu\phi/M^4\gtrsim 1$ while quantum corrections are kept under control and the irrelevant operators in the classical action are technically natural (see e.g. \cite{deRham:2014wfa,Brax:2016jjt} for explicit derivations of these statements).Thus, the non-perturbative classical solutions have a chance of surviving and we can trust their predictions. Similarly, the quantum corrections in non-linear electrodynamics theories can be kept small even in regions where the electromagnetic fields are $\vert F_{\mu\nu}\vert\gtrsim \Lambda^2_{\rm non-linear}$, being Born-Infeld electromagnetism a paradigmatic example.}
 
 An important concern with RBG theories is that the irrelevant operators generated after integrating out the non-dynamical connection are somewhat universal in the sense that they will involve all fields in the matter sector and, at the very least, those of the Standard Model. This universal nature of the RBG theories will typically lead to an EFT in the Einstein frame without any underlying symmetry or structure guaranteeing any non-renormalisation result or naturalness of the resulting interactions like in the case of pure K-essence or non-linear electromagnetisms. Quite the opposite, the very presence of the Higgs field and its potential already points towards the impossibility of having non-perturbative classical solutions based on irrelevant operators without going beyond the regime of validity of the would-be EFT. {Leaving aside these quantum corrections issues, one should not forget that the interest on these theories stems from their non-perturbative properties as classical field theories, as they accommodate a plethora of exact solutions with interesting properties which can expand our dictionary of viable spacetimes. For instance, they are furnished with singularity free cosmological and spherically symmetric backgrounds \cite{Barragan:2009sq,Banados:2010ix,Olmo:2013gqa,Maso-Ferrando:2021ngp,Benisty:2021laq}, as well as wormholes and other compact objects which behave in interesting and unexpected ways (see {\it e.g.} \cite{Lobo:2013adx,Lobo:2013vga,Lobo:2014fma,Lobo:2014zla,Lobo:2020vqh}) which are worth to be explored.}\\

{\bf Acknowledgments.}

JBJ acknowledges support from the ``Atracci\'on  del  Talento  Cient\'ifico'' en Salamanca programme and from project PGC2018-096038-B-I00 by Spanish Ministerio de Ciencia, Innovaci\'on y Universidades. AD  is  supported  by a  PhD contract  with  reference  FPU15/05406  (Spanish  Ministry  ofEconomy and Competitiveness). GJO acknowledges support from the Spanish Agencia Estatal de Investigaci\'on (FIS2017-84440-C2-1-P), by the Generalitat Valenciana (PROMETEO/2020/079), and by the Spanish Research Council (i-COOPB20462, CSIC). This work has further been supported by the European Union's Horizon 2020 Research and Innovation (RISE) programme H2020-MSCA-RISE-2017  Grant  No. FunFiCO-777740.

\bibliography{BibRBGEFT}

\end{document}